\documentclass{llncs}
\usepackage{mathptmx}       % selects Times Roman as basic font
\usepackage{helvet}         % selects Helvetica as sans-serif font
\usepackage{courier}        % selects Courier as typewriter font
\usepackage{type1cm}        % activate if the above 3 fonts are not available on your system
\usepackage{makeidx}         % allows index generation
\usepackage{graphicx}        % standard LaTeX graphics tool when including figure files
\usepackage{multicol}        % used for the two-column index
\usepackage[bottom]{footmisc}% places footnotes at page bottom
\usepackage{subfigure}
\usepackage{amsfonts}
\usepackage[cmex10]{amsmath}

\begin{document}

\title{Non-standard memory models with indexed retrieval}
\titlerunning{Indexed Retrieval Model for Neural Networks}  % abbreviated title (for running head)
\author{Gabriele Scheler\inst{1} 
\and 
Martin L Schumann\inst{2}
\and
Johann Schumann\inst{1}}
\authorrunning{Scheler Schumann et al.} % abbreviated author list (for running head)

\institute{Carl Correns Foundation for Mathematical Biology, Mountain View, CA\\
\email{\tt gscheler@gmail.com},\\ 
\texttt{http://www.theoretical-biology.org}
\and
LRZ Muenchen, Germany}
\maketitle    

\section{Introduction}
The standard memory models for neural networks are variants of the
Hopfield network, where feature representations are stored as vectors
in a matrix. Retrieval happens based on similarity between an input
vector and the set of stored vectors in a content-addressable manner
such that the network evolves towards the closest stored attractor. In
other words, the useful property of addressing items in memory
directly by index is lost in Hopfield-style neural network models
(“associative memory”). Here we present a new model which is
extremely simple, derived from biological observation, yet introduces
a significant conceptual advance and technical benefits

%
%\section{Neuron-centric Implementation}
%
The goal was to establish adaptivity based on the neuron-centric hypothesis:
Plasticity is organized by the neuron which regulates its own synapses.
Accordingly we implemented a localist, neuron-centric one-shot learning
method and applied it to a simple pattern classification problem (MNIST). We
were looking for the existence of high information neurons, to act as indices
into the representations. The idea was that we would be able to restore full patterns by indexed retrieval, instead of associative vector retrieval from attractors.

\section{Methods and Results}
\begin{figure}[htb]
\centering
\includegraphics[width=0.5\textwidth]{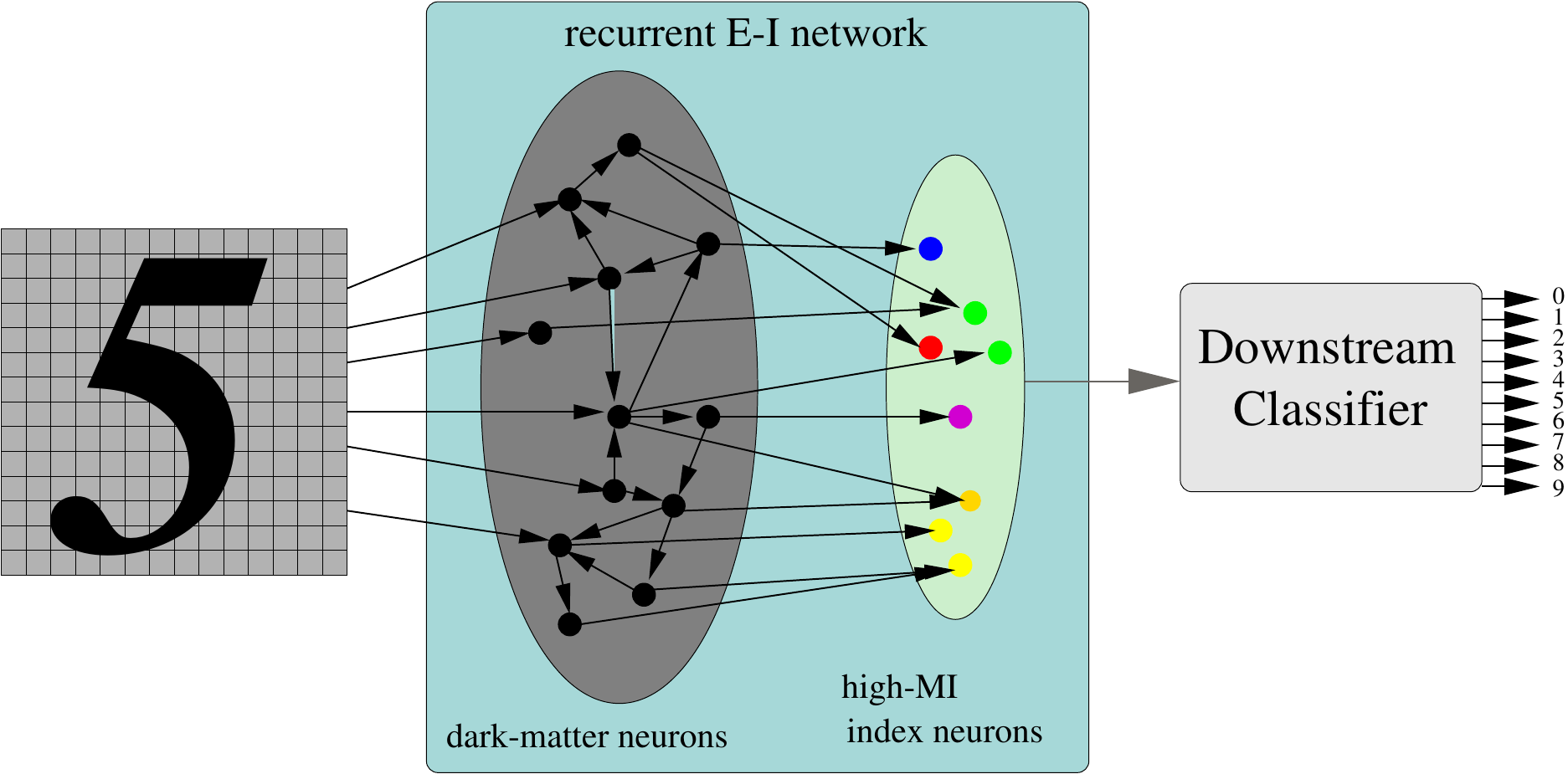}
\caption{Architecture overview: Pattern
input, recurrent E-I network with dark matter neurons and bank of high-MI (index) neurons. The downstream classifier is separate.}
\label{fig:fig1}       % Give a unique label
\vspace{-0.4cm}
\end{figure}

We set up a recurrent excitatory (E) - inhibitory (I) network with a
pattern input area P for digit pattern input (Fig.~\ref{fig:fig1}). 

Input patterns are loaded
into the network and representations develop by recurrent processing overabout 300ms. A downstream classifier was trained to assign the correct categories to patterns. We calculated the mutual information (MI) between patterns and representation for each neuron to find the
neurons with highest MI for each category. Then we used a one-shot neuron-centric plasticity rule, in order to adapt only the highest MI neurons and their connections to E and I neurons (localist plasticity). In this way we derive a
pattern-adapted ('trained') network. Now we have a network that is trained for digit classification. We can stimulate only the highest MI neurons, which we can call index or symbolic neurons, and observe the unfolding of a full feature-based representation on the
network. This allows symbol-based computations derived from feature vectors.  The success of this method is measured by applying the trained classifier to the recalled ('unfolded') representation.
\begin{figure}[tb]
\centering
\includegraphics[width=0.5\textwidth]{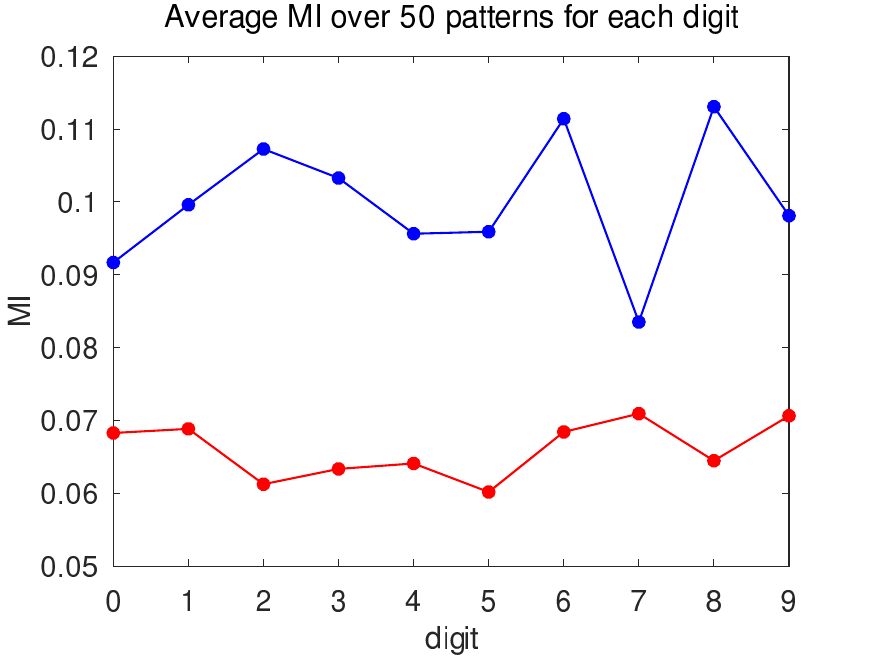}
\caption{Mutual information (MI) between digit representation, summed over the E network activations. Red is before plasticity, blue is after plasticity. MI over the whole network increases for all patterns after learning.}
\label{fig:fig2}       % Give a unique label
\vspace{-0.4cm}
\end{figure}
% Fig.~\ref{fig:fig3}). 
\begin{figure}[htb]
\centering
\ \\[-4ex]
\includegraphics[width=0.7\textwidth]{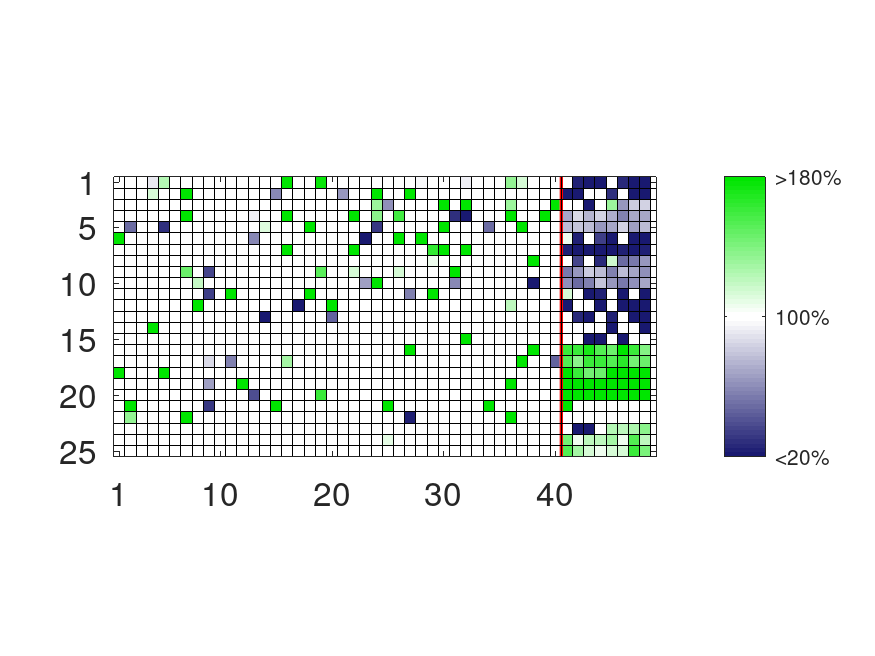}
\ \\[-4ex]
\caption{Difference in activation values before/after learning: mostly unchanged for 1000 E neurons, strong selection by 200 I neurons}
\label{fig:fig4}       % Give a unique label
\vspace{-0.4cm}
\end{figure}

For neurons selected by their high mutual information which code for a pattern (index neurons), all of their input and output synapses are adapted, including the inhibitory connections (localist plasticity). The selection of the correct pattern is guided by inhibition. It is apparent that MI increases after training for each of the patterns learned (Fig.~\ref{fig:fig2}).

As a result, we obtain an implicit hierarchical encoding scheme, consisting of a few high information ‘index’ neurons and many low-level feature neurons. The localist plasticity rule hardly affects activation rates in the network (Fig.~\ref{fig:fig4}), keeping the network properties stable.

\begin{figure}[htb]
\centering
\includegraphics[width=0.3\textwidth]{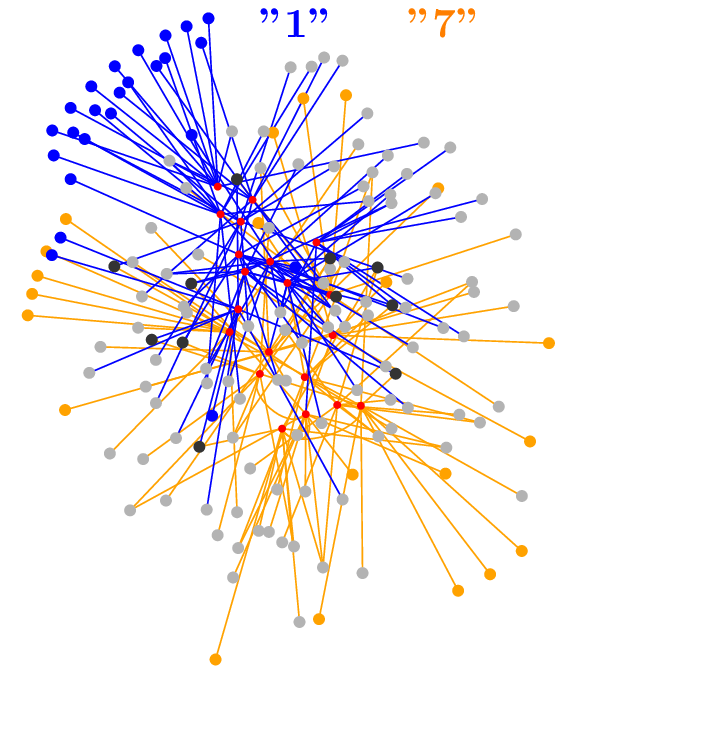}
\includegraphics[width=0.3\textwidth]{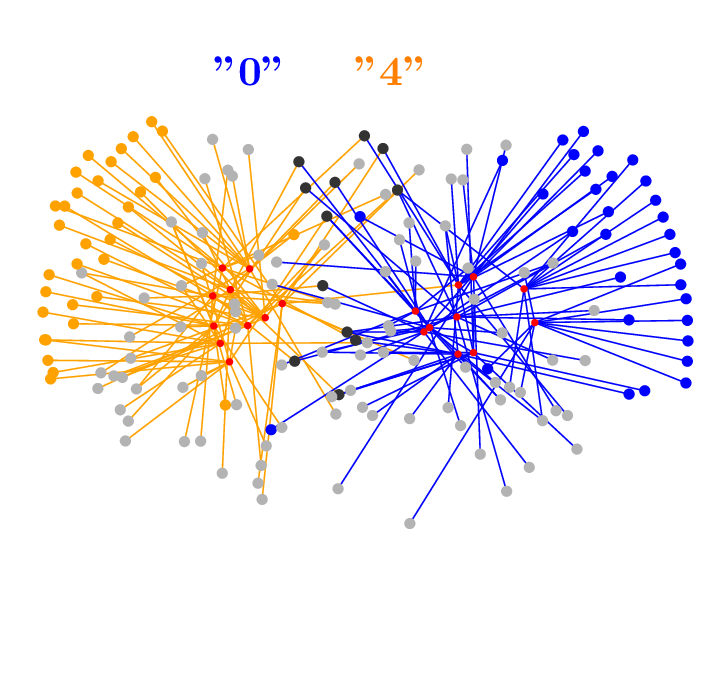}
\includegraphics[width=0.3\textwidth]{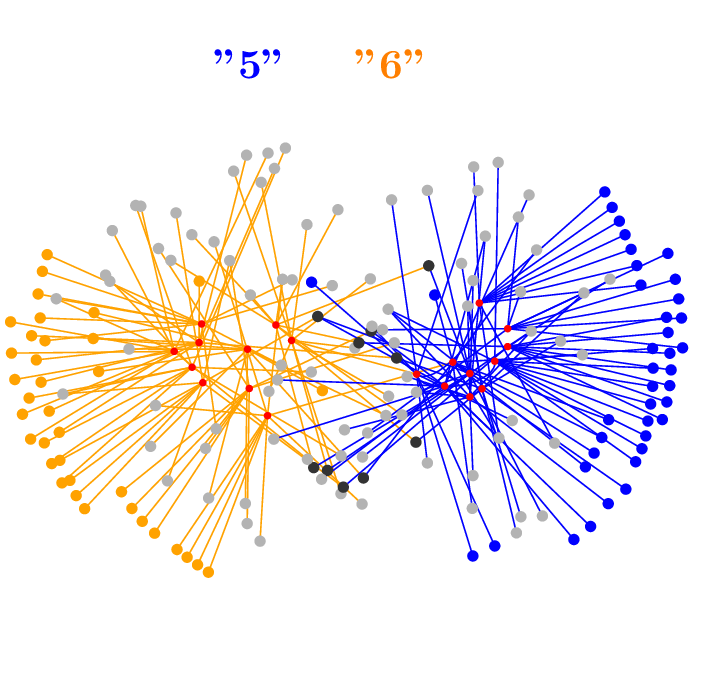}
\caption{Hierarchical, star-shaped pattern representations for digits obtained from neuron-centric plasticity of selected
index neurons. Patterns show more or less overlap in representation space.}
\label{fig:fig5}       % Give a unique label
\vspace{-0.4cm}
\end{figure}

The capacity of a network to classify patterns is instantiated here by 1000
E/200 I neurons for 500 patterns in 10 digits (=pattern classes). Fig.~\ref{fig:fig5} shows the star-shaped representations that develop for several of the classes. 

\begin{figure}[htb]
\centering
{\bf A}
\includegraphics[width=0.4\textwidth]{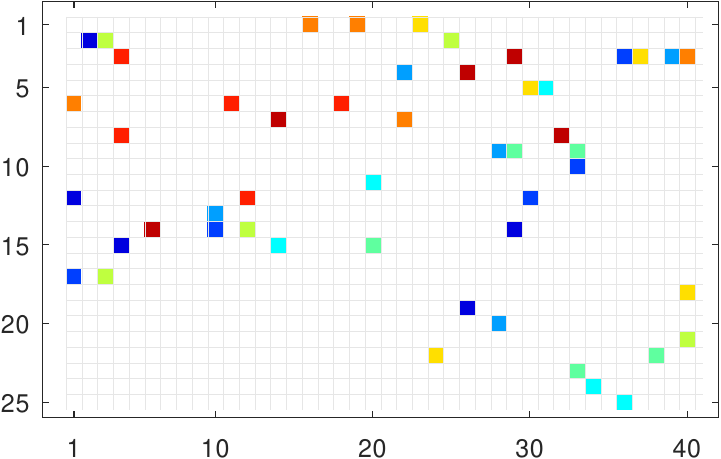}
\hfill
{\bf B}
\includegraphics[width=0.4\textwidth]{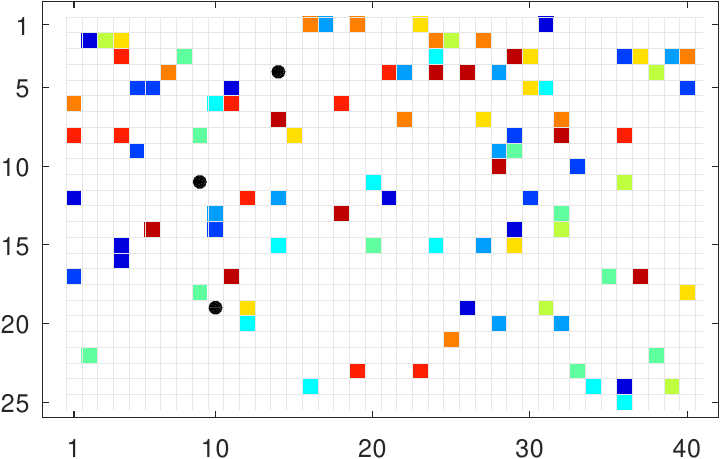}
\caption{{\bf A}: Selection of m=5 index neurons for 10
patterns. There are no overlaps between
indices.
{\bf B}: Selection of m=10 index neurons for
10 patterns. Black spots indicate overlap
between indices for patterns.}
\label{fig:fig6}       % Give a unique label
\vspace{-0.4cm}
\end{figure}

In Fig.~\ref{fig:fig6}, index neurons for each of
the 10 patterns are shown. For m=5, m=10, there are no or almost no overlaps.
For m=20 indices per pattern, there are many overlaps and performance goes
down (not shown). Both m=5 and m=10 are good choices for this
problem. 

\begin{figure}[htb]
\centering
\includegraphics[width=0.5\textwidth]{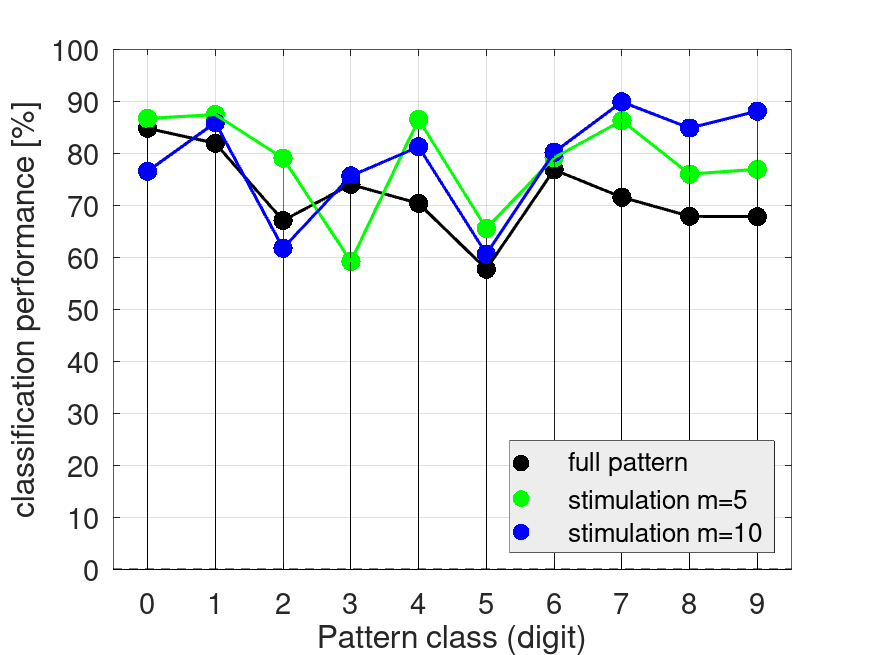}
\caption{Classification performance for stimulating m=5 (green), m=10 (blue) index neurons vs.
learned pattern classifier (full representation, black line)}
\label{fig:fig7}       % Give a unique label
\vspace{-0.4cm}
\end{figure}

The decisive question is the indexed retrieval. Is it possible to unfold recognizable representations on the complete network through stimulation of a few index neurons? Fig.~\ref{fig:fig7} shows the performance when stimulating index neurons only, i.e. 5 or 10 neurons (out of 1000) for each digit pattern (blue, green) compared to stimulation with the full input pattern after plasticity (black line). The performance is comparable for indexed stimulation. 

\section{Summary}
Neuron-centric local plasticity imposes a structure on the network
where index neurons activate the full feature content over their
environment.This is well-suited for crisp and precise symbolic
computation. Patterns can be addressed directly by activating their
indices. In this way, "index" or "symbol" neurons can also function as
control neurons in a complex network setting, interacting with one
another to form a high-level network with access to their associated
feature neurons. This model has been designed with a high degree of biological
realism and a correspondence to brain networks is implied.

%=======================================================
\bibliographystyle{splncs03} % We choose the "plain" reference style
\section*{References}

\begin{itemize}
\item
[1] Martinelli, F.~et al: Flat Channels to Infinity in Neural Loss Landscapes.\\ arXiv:2506.14951
\item
[2] Lin, H (2024): Exploring neural network landscapes: Star-shaped and geodesic connectivity. arXiv:2404.06391
\item
[3] Sonthalia, A.(2024): Do deep neural network solutions form a star domain?\\ arXiv:2403.07968.
\item
[4] Scheler, G etal. (2025): Localist neural plasticity identified by mutual information. J Comput Neurosci. 2025 Jun;53(2):321-331. doi: 10.1007/s10827-025-00901-w
\end{itemize}

\end{document}